\documentclass[journal=jacsat,manuscript=article]{achemso}

\usepackage[version=3]{mhchem} 
\usepackage{amsmath}

\author{Chaudhary Eksha Rani}
\author{Rahul Chand}
\author{Ashutosh Shukla}
\affiliation[Unknown University]
{Department of Physics, Indian Institute of Science Education and Research (IISER) Pune, Pune 411008, India}
\author{G V Pavan Kumar}
\email{pavan@iiserpune.ac.in}
\affiliation[Unknown University]
{Department of Physics, Indian Institute of Science Education and Research (IISER) Pune, Pune 411008, India}

\title[An \textsf{achemso} demo]
  {Evanescent Optothermoelectric Trapping: Deeper Potentials at a Largescale}

\abbreviations{IR,NMR,UV}
\keywords{American Chemical Society, \LaTeX}

\begin{document}

\begin{tocentry}

\begin{center}
    \includegraphics[width = 235 pt]{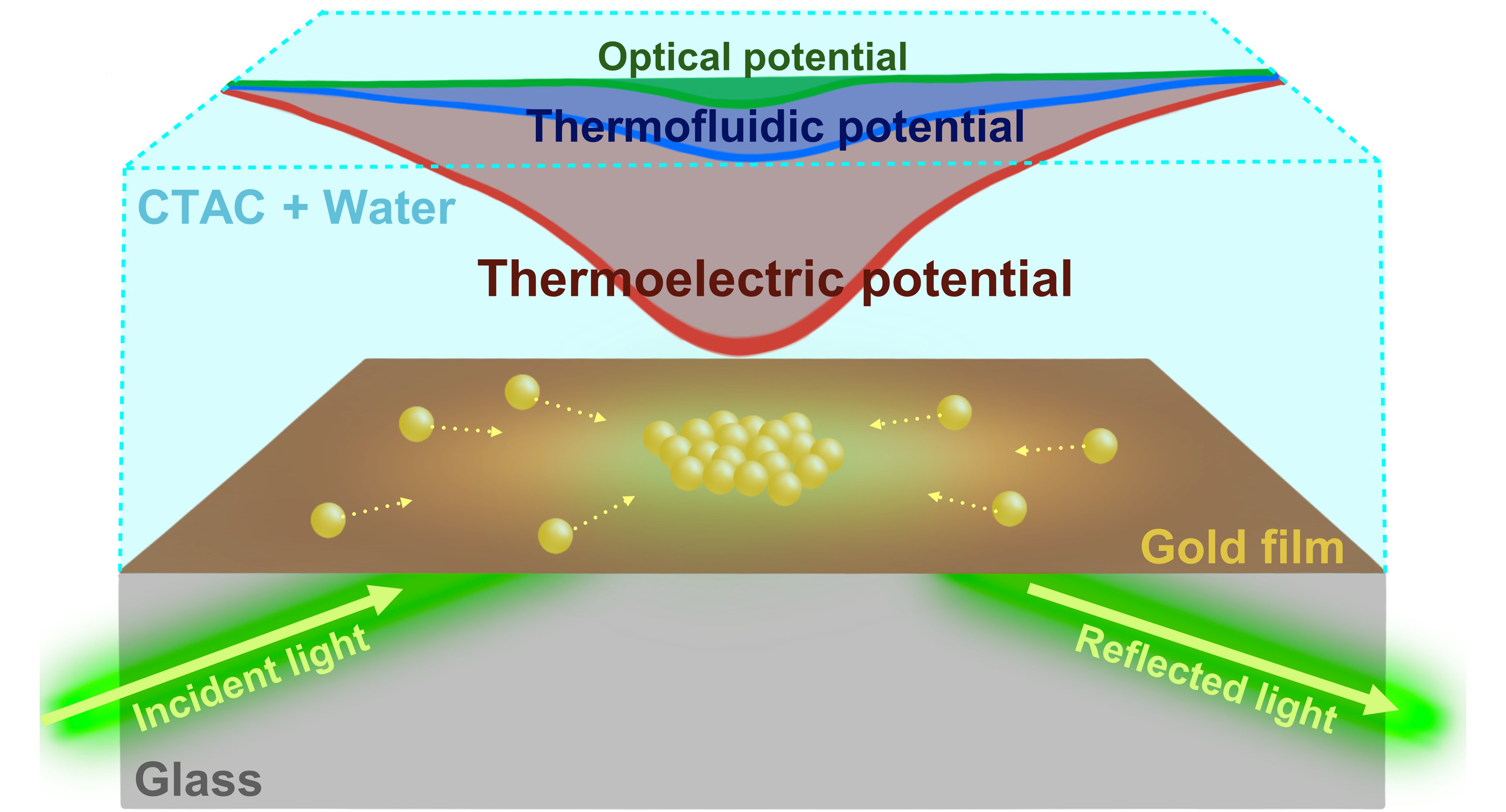}
    \label{For Table of Contents Only}
\end{center}





\end{tocentry}

\begin{abstract}
  Surface plasmons (SP) and their mediated effects have been widely used to manipulate micro- and nanoscale objects of dielectric and metallic nature. In this work, we show how SP excitation can be used to induce thermofluidic and thermoelectric effects to manipulate colloidal dynamics on a large scale. In an evanescent plasmonic trap, temperature gradients induce fluid flow that can facilitate particle accumulation. However, large out-of-plane flows expel particles from the trap, resulting in a shallow trap potential. Here, we numerically demonstrate how adding thermoelectric fields can overpower the optical and hydrodynamic forces to achieve a stable nanoparticle assembly at low excitation powers. We calculate the corresponding optical, fluidic, and thermoelectric trapping forces and potentials. These potentials can be enabled without resonant SP excitation, which requires careful optical alignment. Thus, we explain the mechanism of how, despite weak optical intensities and forces, sufficient trapping force can be supplied via the evanescent optothermoelectric trap to obtain large-scale reversible nanoparticle assemblies, irrespective of their shape, size, or material.
\end{abstract}

\section{Introduction}
Trapping and tweezing of Brownian objects received impetus thanks to the pioneering work of Ashkin and coworkers \cite{ashkin1970acceleration, ashkinObservationSinglebeamGradient1986, ashkinOpticalTrappingManipulation1987a, ashkin1987optical, vsiler2006optical}. The initial and subsequent optical tweezer designs use high numerical aperture objective lenses to spatiotemporally trap and move micro- and nanoscopic objects.\cite{grierRevolutionOpticalManipulation2003a, bradacNanoscaleOpticalTrapping2018, crozierQuoVadisPlasmonic2019, volpeRoadmapOpticalTweezers2023a,nalupurackal2023simultaneous,panja2024nonlinear} In certain applications like microfluidics, biomedicine, and optically driven soft and active matter large-area spatiotemporal traps are required. This has motivated research on ‘macro’ optical traps extending up to a few millimetres. One way to do this is to utilize evanescent waves from the total internal reflection (TIR) of a laser that can confine objects to an area defined by the projected evanescent waves.\cite{righiniLightinducedManipulationSurface2008, yoonOpticalTrappingColloidal2010, yu2007manipulation, luHelicityPolarizationGradient2023} However, creating such large-area optical traps is a challenge. They require large powers of hundreds of milliwatts, multiple laser channels in case of counterpropagating evanescent waves, and careful optical alignment, making them difficult to deploy in power-sensitive environments such as cell manipulation. Additionally, since optical forces are not long-ranged, these traps rely on particle diffusion, making them time-intensive \cite{sasaki1997three}. \\

\noindent An alternate strategy is to utilize these evanescent waves to excite surface plasmons (SP) \cite{barnes2003surface}, facilitating radiative and absorptive channels for energy exchange that lead to opto-thermal potentials \cite{zhangPlasmonicTweezersNanoscale2021a, kang2015trapping, kotsifaki2019plasmonic, juan2011plasmon, quidant2008surface}. Our group has developed large-area optothermal trapping and tweezing methods that can be harnessed for single-molecule spectroscopy and dynamic lithography \cite{patra2014plasmofluidic, patra2016large}. However, these traps depend on the size, shape, and material of the particle, making them highly selective \cite{patra2014plasmofluidic}. These trapping methods are also extremely sensitive to the plasmonic resonances of the metal, which requires tuning the incident wavelength \cite{cuche2013sorting}, and polarisation \cite{sharma2021optothermal}. Moreover, SP need to be excited resonantly at the surface plasmon resonance (SPR) angle to maximize the optical forces, which requires additional experimental equipment and optical alignment \cite{garces2006extended, volpe2006surface}. An unavoidable consequence of SP excitation is the large temperature gradients, which induce strong fluid flows that can make the trap unstable. It is imperative to develop optical trapping mechanisms that utilise the consequent effects of heat and fluid flow while relaxing the resonant excitation conditions to achieve large-scale, stable traps.\cite{kotsifakiRoleTemperatureInducedEffects2022a, chen2021heat, kollipara2023optical, kotnala2019overcoming, sharma2020large, peng2020opto} Here, we demonstrate one such mechanism using the opto-thermoelectric (OTE) effect, which enables large-scale reversible trapping using low power non-resonant optical excitation irrespective of the particle characteristics.\cite{shuklaOptothermoelectricTrappingFluorescent2023, wangGrapheneBasedOptoThermoelectricTweezers2022b, liuNanoradiatorMediatedDeterministicOptoThermoelectric2018c, liOptothermallyAssembledNanostructures2021a, kotnalaOptothermoelectricSpeckleTweezers2020c, kolliparaThermoElectroMechanicsIndividualParticles2019d} While most OTE-based trapping studies have focused on small-area traps, a systematic study on large-area OTE traps is necessary. \\

\begin{figure}[!ht]
\centering
\includegraphics[width=\textwidth]{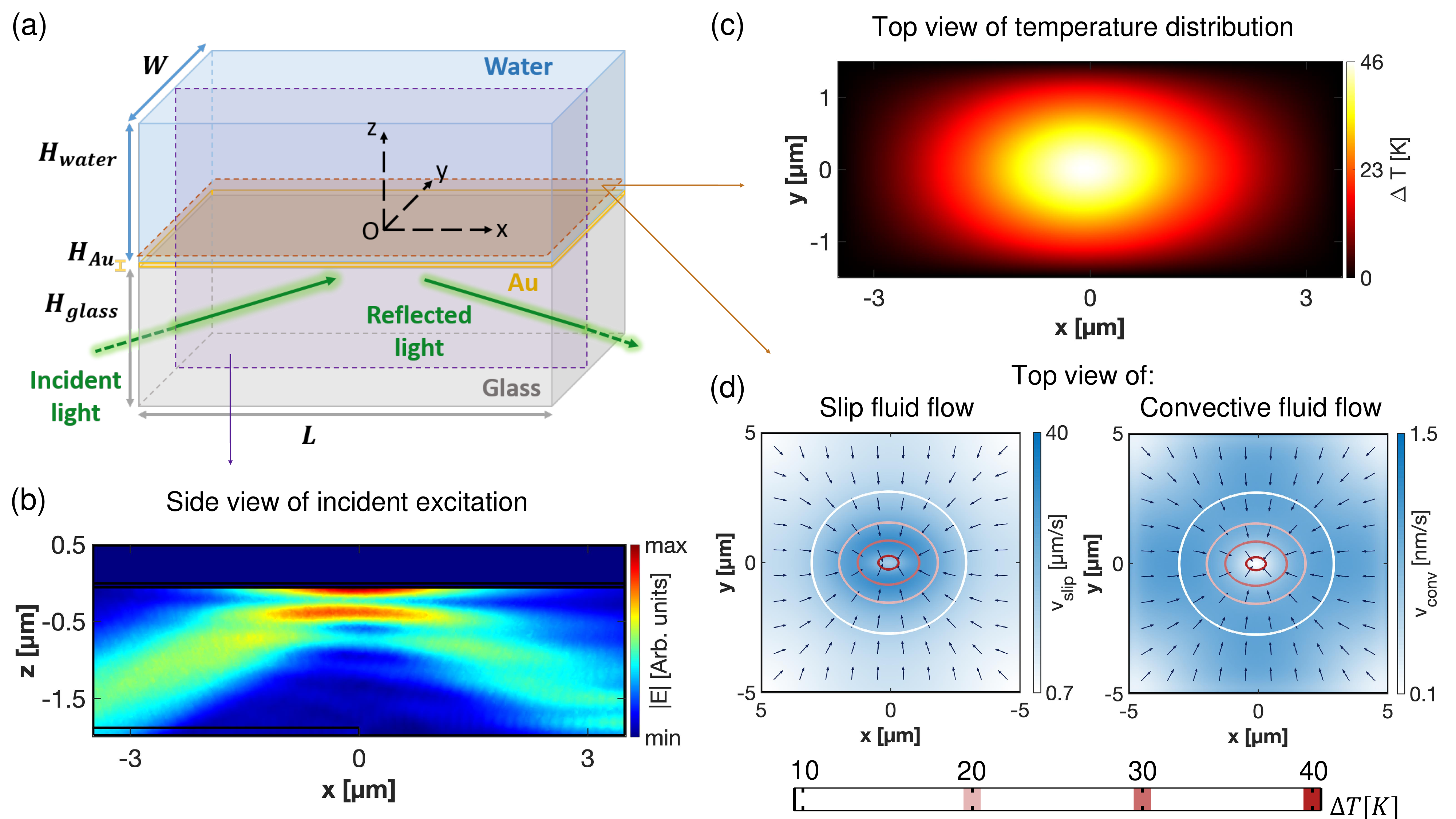}
\caption{(a) Geometry simulated to compute the fields. (b) XZ profile of Gaussian wave exhibiting TIR at an angle $\theta_i=67^o$, greater than the glass-water critical angle $\theta_c=66.5^o$. (c) XY profile of elliptical temperature distribution on the Au film (thermal conductivity 150 W/m.K) due to SP excitation. (d) Laser heating induced slip and convective fluid flows in the XY plane 100 nm above Au film in an enlarged geometry.}
\label{fig1}
\end{figure}

\noindent In this article, we present our numerical study of evanescent OTE traps that unveil the relative contributions of optical, hydrodynamic, and OTE fields, showing a boost in trapping forces when OTE is employed. Consider the geometry as shown in Figure \ref{fig1}(a), consisting of the glass and water domains with a 50 nm thin film of gold (Au) sandwiched in between. Gold nanoparticles (AuNP) of diameter $2r = 400$ nm, 200 nm, 300 nm, or 100 nm are considered diffusing in water. SP are evanescently excited in the Au film that exert negligible optical force. SP also lead to a temperature hotspot in the excitation region that induces a fluid flow directed towards the hotspot. The in-plane flows exert strong hydrodynamic forces to assemble AuNPs, but the out-of-plane flow tends to make the thermofluidic trap unstable. To stabilize the trap, we introduce a thermoelectric field by considering the presence of ions in the fluid, which migrate under a temperature gradient. Since the thermoelectric field is generated by optically-created temperatures, the resultant force is called optothermoelectric force. It overpowers the thermofluidic trapping forces by multiple orders of magnitude and enables a stable assembly of AuNPs. Finite element method-based simulations in COMSOL multiphysics are used to compute the plasmonic optical fields, temperature distribution, optothermally induced fluid flow, and thermoelectric fields. The corresponding forces and potentials experienced by an AuNP are calculated to explain the relative role played by each. The material parameters considered are directly taken from the in-built material library of COMSOL multiphysics unless mentioned otherwise. \\

\section{Model and theory}

\noindent \textbf{Electromagnetic field:} Figure \ref{fig1} (a) shows the Kretschmann configuration used to excite surface plasmons in the gold film \cite{patra2014plasmofluidic}. A p-polarised Gaussian electromagnetic wave is incident on the glass-water interface at an angle $\theta_i = 67^o$, such that it undergoes total internal reflection (critical angle $\theta_c \sim 66.5^o$). Consequently, evanescent waves are generated in the water (rarer) medium, which can excite SP in the metal film. 
The SP intensity can be controlled by changing the incident power or polarisation \cite{maier2007plasmonics} since only the TM electromagnetic wave can excite SPs in this configuration. The electromagnetic field is simulated using the ‘electromagnetic wave frequency domain’ (\textit{ewfd}) module. Being computationally expensive, \textit{ewfd} limits the maximum geometrical dimension that can be simulated. Employing scattering boundary conditions, the geometric dimensions are carefully chosen to avoid any boundary reflections from the reflected wave: $H_{water} = 0.5$ $\mathrm{\mu}$m, $H_{glass} = 2$ $\mathrm{\mu}$m, $L = 7$ $\mathrm{\mu}$m, $W = 3$ $\mathrm{\mu}$m. The XZ profile of the incident light (p-polarised, $\lambda = 532$ nm, power 10 mW) undergoing TIR is shown in Figure \ref{fig1}(b), which gives surface fields at the Au-water interface. It should be mentioned that such surface fields are not observed when s-polarized incident light is simulated, which confirms the SP excitation. \\

\noindent \textbf{Thermofluidic field:} SP generation in the metal film leads to energy dissipation via the Joule effect \cite{baffouThermoplasmonicsHeatingMetal2017b}. In the temperature homogenization regime, the temperature distribution will directly correspond with the intensity distribution. An obliquely incident Gaussian beam will result in an elliptical intensity spot \cite{wang2009propulsion} which corresponds to an elliptical temperature distribution on the Au surface, as shown in Figure \ref{fig1}(c). The temperature magnitude can be controlled using the SP intensity by tuning the power or polarization of the incident light. Ideally, s-polarised light will not lead to any temperature rise due to its inability to excite SP. The SP-mediated heat is modeled in COMSOL using the ‘heat transfer in solids and fluids’ (\textit{ht}) module coupled with the \textit{ewfd} module. \\
The heated Au film leads to temperature-induced fluid flows of two types: (1) convective fluid flow arising from temperature-dependent density of water and (2) slip fluid flow arising from temperature-dependent interfacial interactions \cite{bregulla2016thermo}. These fluid flows are simulated using the ‘laminar flow’ (\textit{spf}) module coupled to the \textit{ht} module. The Au-water interface is set to slip boundary condition with a dimensionless slip coefficient $\sigma_T = 0.001$ \cite{franzl2022hydrodynamic} to compute the slip flow and no-slip boundary condition to compute the convective flow. All other walls are set to open boundary conditions. However, experiments with colloidal solutions are usually performed in larger fluid columns to minimize boundary effects on colloidal diffusion. But, since computing such large domains is extremely memory-expensive for \textit{ewfd}, we used an alternative heat source called ‘deposited beam power’ that models the temperature profile of a laser-heated surface without explicitly modeling the laser. Thus, eliminating \textit{ewfd}, we enlarged the fluid column while significantly reducing the computational resources needed: $H_{water} = 5 $ $\mathrm{\mu}$m, $H_{glass} = 2$ $\mathrm{\mu}$m, $L = 10$ $\mathrm{\mu}$m, $W = 10$ $\mathrm{\mu}$m. The parameters of the Gaussian laser used by ‘deposited beam power’ are carefully chosen to match the temperature profile generated via the \textit{ewfd}-mediated heat source. The corresponding slip and convective fluid flow in the XY plane (100 nm above Au surface) are shown in Figure \ref{fig1}(d), where the contours represent the temperature rise against a room temperature of 298 K. \\

\noindent \textbf{Thermoelectric field:} If the fluid contains ions that migrate differently under a temperature gradient, then spatial separation of ions can be achieved. \cite{kolliparaThermoElectroMechanicsIndividualParticles2019d, duhrWhyMoleculesMove2006} This will produce an electrostatic field in the fluid called the thermoelectric field $E_T=\frac{k_B T\nabla T}{e}\frac{\sum_i Z_i c_i S_{T_i}}{\sum_i Z_i^2 c_i}$, directed along the temperature gradient. Here, $i$ indicates the ionic species, $k_B$ is the Boltzmann constant, $e$ is the elementary charge, $Z$ is the charge number, and $c$ is the concentration of ions. Once $E_T$ is activated, it can exert thermoelectric force $F^{OTE} = Q\cdot E_T$ on any particle having total charge $Q$, driving it towards or away from the hotspot. The temperature $T$ and its gradient $\nabla T$ are extracted from the COMSOL model to compute the thermoelectric fields and forces.

\section{Results and discussion}

\noindent \textbf{Optical force:} A spherical AuNP is modeled in the water domain 20 nm above the Au surface (which is of the order of Debye length) \cite{quidant2007optical, wang2009propulsion, min2013focused} at the center of the SP excitation spot. The optical forces exerted due to SP fields are calculated by integrating the Maxwell stress tensor over the AuNP surface \cite{jonesOpticalTweezersPrinciples2015a}. The resultant optical forces are of the order $10^{-22}-10^{-20}$ N , which is negligibly small compared to the Brownian forces experienced by these particles in the range $10^{-16}-10^{-14}$ N  \cite{cuche2013sorting}. Such weak optical forces are due to the non-resonant excitation of SP at $\theta_i$, where only a small fraction of incident power is coupled to SP, and the majority is reflected back into glass. To maximize the optical forces up to a femtonewton order, (1) SP can be resonantly excited at SPR angle well beyond $\theta_c$ \cite{garces2006extended, volpe2006surface}, or (2) laser power can be increased \cite{garces2006extended}, or (3) particle resonances can be matched with the laser wavelength \cite{cuche2013sorting} and polarization \cite{patra2014plasmofluidic, xu2018direct}. Experimentally, it may not always be possible to tune $\theta_i$ to the SPR angle or optimize the trap to match particle resonances. Moreover, largescale trapping will require large input powers. To execute an efficient trap without changing the excitation angle, power, or wavelength, it is imperative to utilize the dissipated optical energy in the form of heat to manipulate colloids. Given the small magnitude of optical forces, we consider that the SP fields have minimal effect on particle dynamics. \\

\begin{figure}[ht]
\centering
\includegraphics[width=\textwidth]{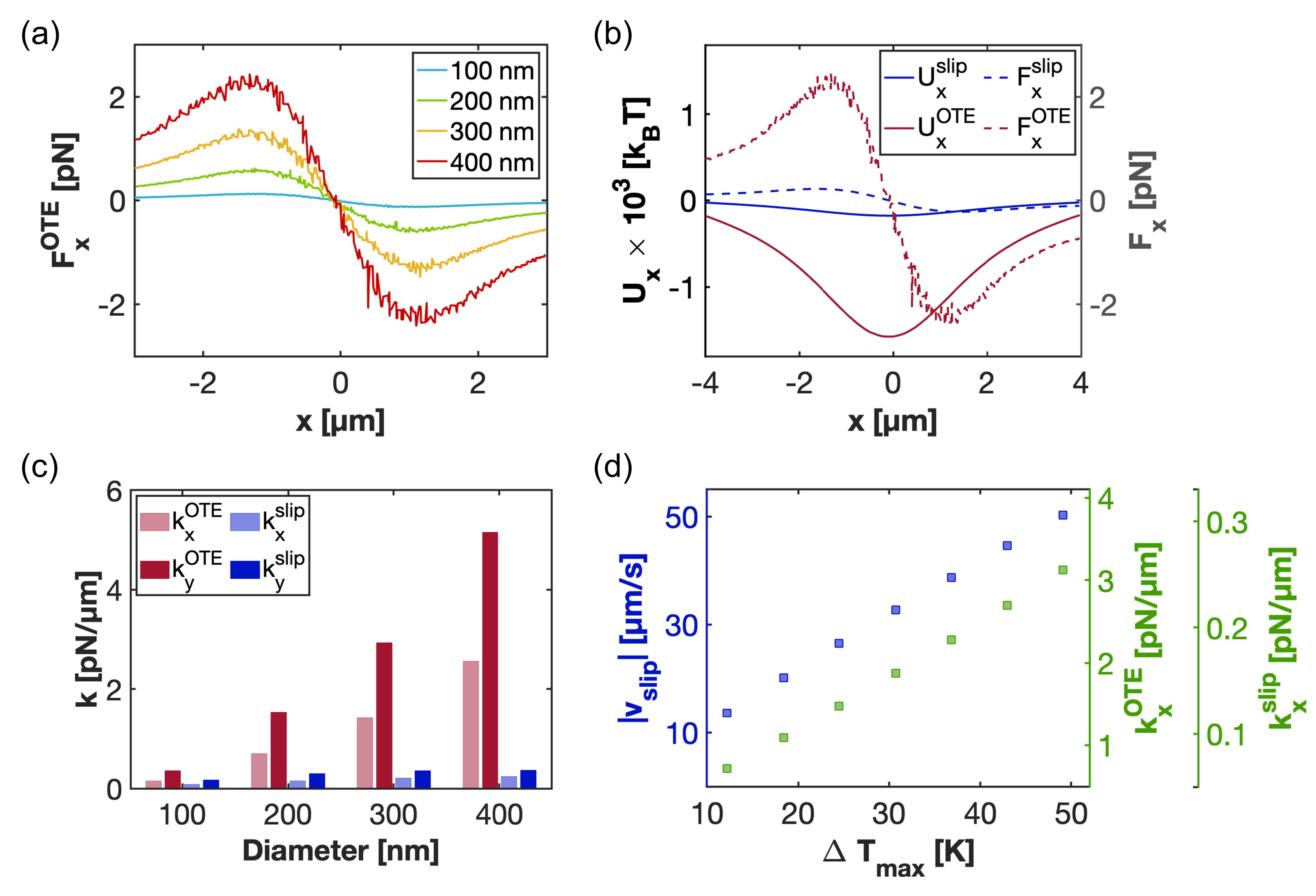}
\caption{(a) x-component of thermoelectric force $F_x^{OTE}$ exerted on AuNPs of different diameters. (b) Comparison between trap potential $U_x$ felt by a 400 nm AuNP along X due to OTE and slip forces. (c) Comparison between the trap stiffness (\textit{k}) along X and Y due to $F_{x,y}^{OTE}$ and $F_{x,y}^{slip}$ forces for AuNP of different diameters. (d) Variation of maximum slip flow velocity and trap stiffness for 400 nm AuNP, due to $F_x^{OTE}$ and $F_x^{slip}$ forces, with maximum temperature rise $\Delta T_{max}$.}
\label{fig2}
\end{figure}

\noindent \textbf{OTE and hydrodynamic slip force calculation:} The most prominent heat-mediated effects widely used in trapping are the thermophoresis\cite{kollipara2023optical}, thermofluidic \cite{chand2023emergence} and thermoelectric \cite{shuklaOptothermoelectricTrappingFluorescent2023} effects. Since metallic colloids do not exhibit thermophoresis, AuNP does not migrate under a temperature gradient \cite{bregulla2016thermo} and will be driven by the fluid flow in the system under the effect of hydrodynamic forces. We have done a numerical study to compare the two remaining effects and found that the thermoelectric force is much stronger and gives greater control over particle dynamics than the thermofluidic forces. To study the thermoelectric effect, consider an AuNP freely diffusing in a 5 mM solution of a surfactant CTAC (cetyltrimethylammonium chloride) that dissociates into CTA$^+$ and Cl$^-$ ions. Above the critical micellar concentration, CTAC self-assembles into micelles, forming macrocations. The CTAC molecules get adsorbed on the AuNP surface, forming a double layer and making it positively charged with a zeta potential $\zeta$. We have approximated $\zeta = 63.6$ mV, 73.3 mV, 76 mV, and 80 mV for $2r = 100$ nm, 200 nm, 300 nm, and 400 nm, respectively \cite{lin2018opto}, considering the general trend of $\zeta$ measurements in 5 mM CTAC. Without a temperature gradient, the positively charged AuNP, CTAC micelles, and Cl$^-$ ions are randomly dispersed in the solution. Under the SP-mediated temperature gradient, the micelles and Cl$^-$ ions will move differently owing to their different Soret coefficients (CTAC micelles: $10^{-2}$, Cl$^-$ ions: $7.18\times 10^{-4}$) \cite{lin2016light}, leading to a spatial separation between positive and negative ions. This generates an electrostatic field $E_T$ that can exert thermoelectric force $F^{OTE} = (4\pi r^2 \sigma)\cdot E_T$ on a spherical particle having surface charge density $\sigma$ \cite{ohsawa1986zeta}. Using $T\nabla T$  from the COMSOL simulation, we computed the force components  $F_x^{OTE}$, $F_y^{OTE}$, and $F_z^{OTE}$ exerted on an AuNP located 20 nm above the Au surface. $F_x^{OTE}$,  shown in Figure \ref{fig2}(a), in the pN order, is much stronger than the minimum fN order of trapping force required to confine AuNP at the hotspot. The magnitude and slope increase with particle size $2r$  due to the greater zeta potential of larger particles, giving them greater surface charge. \\

\noindent To compare with the thermofluidic effects, we estimated the hydrodynamic force on AuNP due to temperature-induced fluid flows. For a spherical particle with a small Reynolds number ($\sim 10^{-6}$), the hydrodynamic force is given by the Stokes equation $F = \gamma \cdot v = 6\pi \eta r v$, where $\eta$ is the dynamic viscosity of the fluid, $v$ is the fluid velocity, and $\gamma$ is the friction coefficient near a surface, which gets modified compared to its bulk value $\gamma_0$ as $\gamma=\alpha\gamma_0$ \cite{brenner1961slow, jonesOpticalTweezersPrinciples2015a}. Using the slip or convective flow velocity from COMSOL, we can calculate the hydrodynamic slip and convective forces, $F^{slip}$ and $F^{conv}$, respectively.\\

\noindent \textbf{OTE vs hydrodynamic force:} To compare the OTE and the fluid forces, we will consider the 400 nm AuNP. Figure \ref{fig2}(b) shows the x-component of OTE, $F_x^{OTE}$, and hydrodynamic slip forces $F_x^{slip}$, along with the respective potentials $U_x$. We consider only $F_x^{slip}$ for comparison as it is larger of the two fluid forces. While both $F_x^{OTE}$ and $F_x^{slip}$ are strong enough to trap AuNPs, $F_x^{OTE}$ is ten times stronger with a thermoelectric trap stiffness $k_x^{OTE} = 2.56$ pN/$\mu$m compared to the thermofluidic trap stiffness $k_x^{slip} = 0.24$ pN/$\mu$m. This results in a thermoelectric potential $U_x^{OTE}$ that is over eight times deeper than the thermofluidic potential $U_x^{slip}$, trapping the AuNP in the surface plane above the hotspot. Due to the elliptical shape of the temperature distribution, as is evident from the contours in Figure \ref{fig1}(d), the temperature gradients along Y are greater than those along X. This results in steeper force profiles and a larger trap stiffness along Y, $k_y \sim 2k_x$, for both OTE and hydrodynamic slip forces as is shown in Figure \ref{fig2}(c). Thus, AuNP diffusing in water will be driven toward the hotspot collectively by $F_{x,y}^{OTE}$ and $F_{x,y}^{slip}$ (in-plane forces), which facilitate an elliptical region of closely assembled AuNP above the hotspot. It should be emphasized that temperature is the primary control parameter for both thermofluidic and thermoelectric traps, as depicted by Figure \ref{fig2}(d) (for the 400 nm AuNP). The maximum slip velocity $v_{slip}$ varies linearly with $\Delta T_{max}$ due to its dependence on temperature as $\Vec{v}_{slip} = \chi \nabla T/T$\cite{bregulla2016thermo}, directly affecting the thermofluidic trap stiffness $k_x^{slip}$, where, $\chi$ is the thermo-osmotic slip coefficient. The rising temperature will also strengthen $E_T$, leading to a stronger $F^{OTE}$ and increasing $k_x^{OTE}$. Figure \ref{fig2}(d) also shows that $k_x^{slip}$ is one order of magnitude smaller than its thermoelectric counterpart $k_x^{OTE}$. The y-components of the forces and potentials exhibit a similar trend as the x-components but with steeper force profiles and deeper potentials. Note that, up to an extent, $F^{OTE}$ can be increased while keeping $F^{slip}$ constant by increasing the CTAC concentration, enhancing $E_T$ and thus the forces. However, to calculate the precise enhancement, we require zeta potential of AuNP in the increased concentration. For further comparison, we will continue the analysis considering 5 mM concentration of CTAC. \\

\begin{figure}[ht]
\centering
\includegraphics[width=\textwidth]{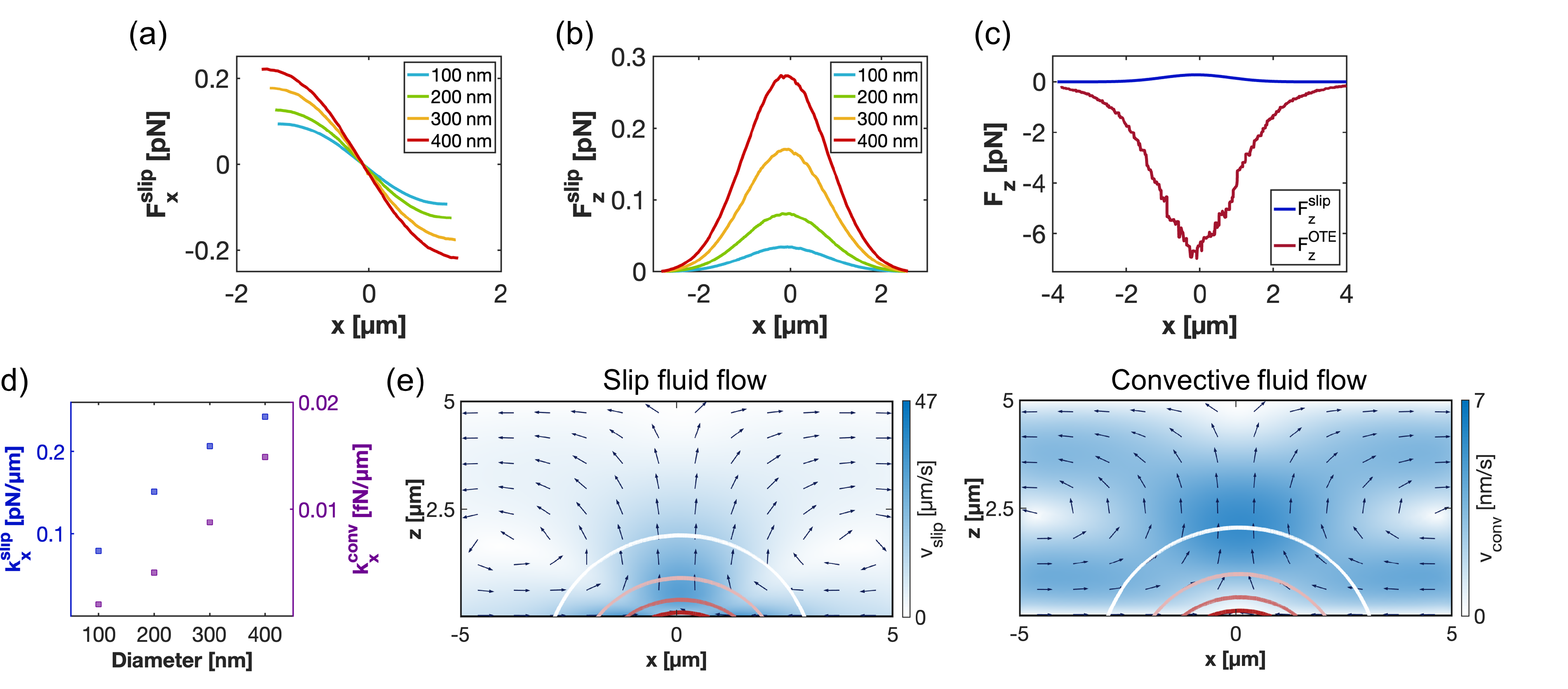}
\caption{Hydrodynamic forces: (a) and (b) represent the $F_x^{slip}$ and $F_z^{slip}$ forces on AuNP of different diameters. (c) Comparison between $F_z^{slip}$ and $F_z^{OTE}$ forces showing their opposing nature. (d) Variation of trap stiffness with AuNP diameter due to hydrodynamic slip and convective forces. (e) XZ profile of slip and convective fluid flow, along with contours showing temperature rise.}
\label{fig3}
\end{figure}

\noindent \textbf{Optical field enhancement:} SP, being evanescent fields, decay exponentially away from the Au surface, which is why certain applications like biosensing and Raman spectroscopy require the objects to be very close to the surface. But, temperature-induced fluid flows like slip flow can pose a challenge. The x- and z-components of hydrodynamic slip forces, $F_x^{slip}$ and $F_z^{slip}$, on an AuNP located 20 nm above the surface are shown in Figures \ref{fig3}(a) and \ref{fig3}(b), respectively. An increase in forces with particle size is expected due to the smaller diffusion coefficient of larger particles and the dependence of hydrodynamic forces on particle size. These sub-pN $F^{slip}$ forces are sufficient to influence the Brownian dynamics of the AuNP and result in a strong enough thermofluidic trap that can assemble AuNP above the hotspot. However, the assembled AuNP are unable to interact with the evanescent SP field due to the slightly stronger out-of-plane force $F_z^{slip}$, which expels the AuNP out of the trap volume. This makes the trap unstable in the vertical dimension. The instability is countered by the negative $F_z^{OTE}$, as plotted in Figure \ref{fig3}(c), which is directed towards the hotspot. It is seven times stronger than the positive $F_z^{slip}$, effectively opposing the upward pull and enabling a stable trap in the vertical dimension. Being closely confined to the surface, the particles can interact with the evanescent SP fields and create subwavelength gaps where the electric field can be localized and enhanced by multiple orders. Such confinement is crucial for applications like plasmonic sensing \cite{tiwari2021single} and multiple scattering experiments like lasing from colloidal assemblies \cite{trivedi2022self}. Inferring from Figure \ref{fig2}(d), large temperatures will tend to strengthen $F_z^{slip}$ fluidic forces, making the trap unstable at large powers. But, in the presence of ions, the temperatures will also reinforce the thermoelectric field and $F_z^{OTE}$ will facilitate stable trapping close to the surface. Hence, evanescent-OTE trap facilitates stable potential at both low powers, when thermofluidic potentials are shallow, and at high powers, when thermofluidic traps are unstable. \\

\noindent \textbf{Hydrodynamic convective forces:} Hydrodynamic convective forces $F^{conv}$ are calculated using the convective flow velocity $v_{conv}$ and are in the order $10^{-2}-10^{-3}$ fN. Figure \ref{fig3}(d) compares the trap stiffness due to $F_x^{conv}$ and $F_x^{slip}$ for different sizes of AuNP. While both $k_x^{conv}$ and $k_x^{slip}$ increase with particle size, $k_x^{conv}$ shows a concave-up increase and $k_x^{slip}$ a concave-down increase. This can be understood from the dependence of the hydrodynamic force on $\gamma$ and $v$. As the particle size increases, its center-to-surface distance increases and $\gamma$ increases for both kinds of flows. However, from Figure \ref{fig3}(e), two distinct characteristics of $v$ can be observed: (1) $v_{slip}$ is three orders of magnitude stronger than $v_{conv}$, and (2) $v_{slip}$ is stronger near the interface while $v_{conv}$ is stronger in the bulk fluid, that is, $v_{slip}$ decreases while $v_{conv}$ increases as we go higher in $z$. This results in a steeper increase in $k_x^{conv}$ compared to a flatter rise in $k_x^{slip}$. Despite such trends, the slip forces result in a much stiffer trap, exceeding the trap stiffness due to convective forces by atleast four orders of magnitude. From this, we can disregard the effect of convective fluid flow on AuNP dynamics. Additionally, for smaller fluid columns like the one considered here, the convective fluid flows are largely suppressed, and slip fluid flow dominates the particle dynamics. \\

\begin{figure}[ht]
\centering
\includegraphics[width=240 pt]{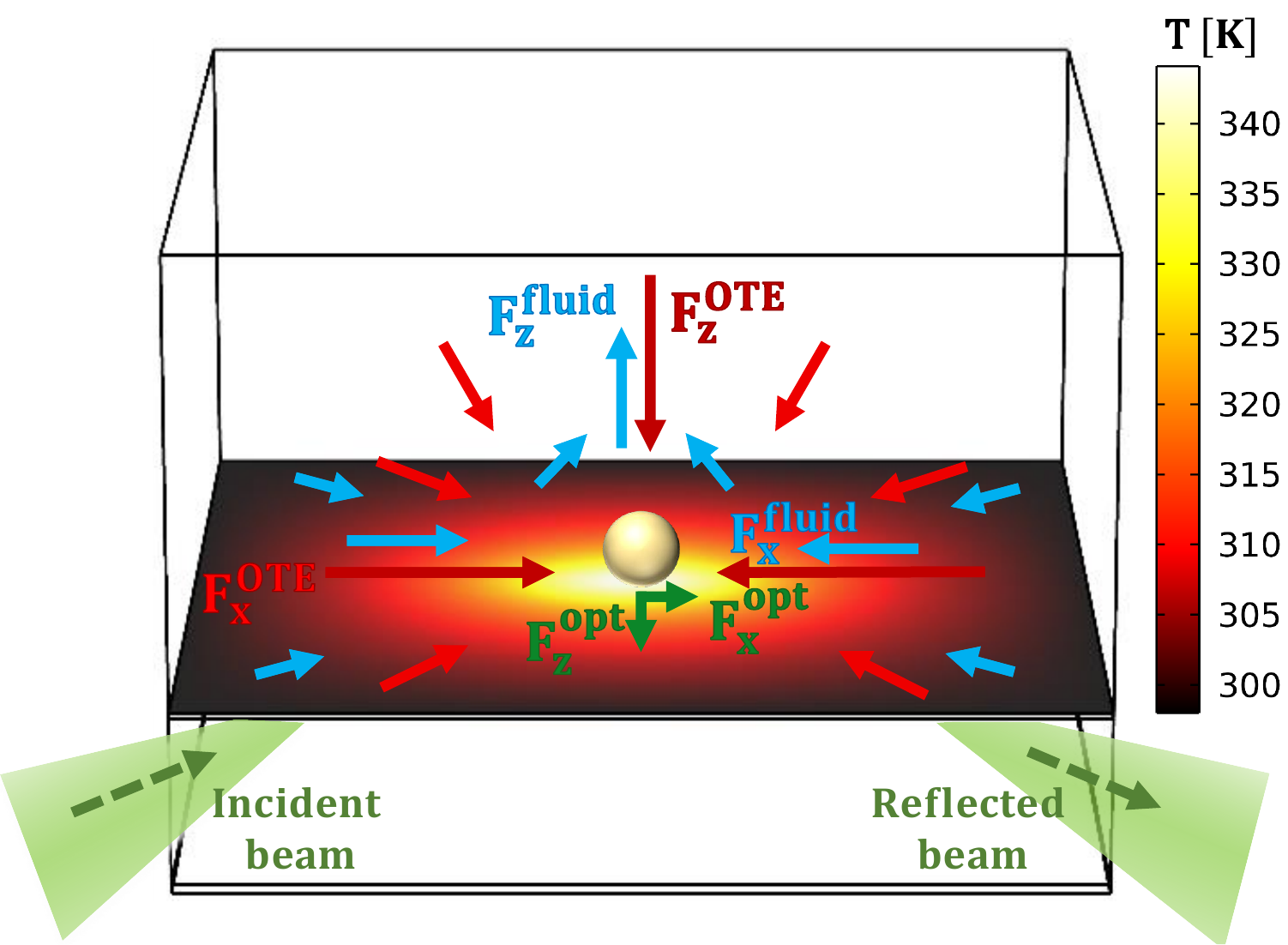}
\caption{Schematic representing the different forces acting on an AuNP suspended above an evanescently excited Au film.}
\label{fig4}
\end{figure}

\noindent \textbf{Control parameters of the trap:} A schematic of the resultant forces is shown in Figure \ref{fig4}, depicting the respective forces acting in the system, mediated by the temperature rise from laser heating in an evanescent-OTE trap. The primary control parameter of both hydrodynamic and thermoelectric forces is the temperature distribution, which depends on the SP intensity. Besides varying SP intensity directly from the laser power, the polarization can be switched between p- and s- to control the fraction of incident light converted to SP. Also, illuminating at SPR angle can induce large temperature rise even at small laser powers. The trap will vanish instantaneously upon the removal of temperature gradients, making the assembly formation a reversible process. Loosely focusing the laser at an oblique angle provides large-scale illumination and temperature gradients over a large area, which dictates the assembly dimension. Once the trap begins to form, the accumulated AuNP will interact with the SP fields, leading to intensity enhancement. This will heat the AuNP, acting as positive feedback that enhances the hydrodynamic and thermoelectric effects. Moreover, thermoelectric forces can additionally be controlled by the ionic concentration, given the direct dependence of the thermoelectric field on concentration.  

\section{Conclusion}

\noindent We have demonstrated the working mechanism of an evanescent-optothermoelectric trap that can reversibly assemble metallic nanoparticles of any shape or material on a large scale, with particle radius down to 50 nm. While the simulations depict results for a smaller dimension, the same can be extended to a larger scale to obtain similar results. Large-scale TIR-illumination can access a large working area spanning hundreds of microns without using complex experimental optics or delicate fabricated structures, giving large-scale trap potentials. The potential depth depends on the temperature magnitude generated due to the illumination and ionic concentration. The temperature magnitude is controlled by incident power, polarization, and excitation angle. Since optical forces are not the major player, the trap is independent of particle characteristics and requires no resonance matching, wavelength tuning, high NA lens, or high powers. While optically mediated hydrodynamic forces lead to a thermofluidic trap, it is unstable in the vertical dimension. Whereas, thermoelectric forces can be 10 times stronger than hydrodynamic forces and facilitate deeper potentials to trap and assemble nanoparticles close to the surface. Such closely assembled particles can achieve excellent electric field enhancement, having direct applications in SERS \cite{patra2014plasmofluidic}. Careful tuning of $F^{OTE}$ and $F^{slip}$ can result in applications like particle sorting without the need for multiple lasers \cite{cuche2013sorting}. Importantly, the large illumination area ($\sim$sub-mm) results in very small intensities (tens of mW in hundreds of $\mu$m$^2$), yet enough to assemble nanoparticles on such a large scale \cite{patra2016large}, compared to 0.1 mW/$\mu$m$^2$ used to assemble over a few microns \cite{lin2016light}. Once the assembly is formed, particles can self-heat, sustaining the trap even without metal film. Such an assembly can potentially be transferred as a single entity \cite{tiwari2021single}. Moreover, dielectric particles can be trapped using this method despite their thermophobic behavior \cite{garces2006extended}. We anticipate large-area traps to be useful in dynamic lithography of small entities such as Brownian colloids and bio-macromolecules.  The strategy can also confine larger biological entities such as cells. This expands the capability of evanescent optical trapping from molecular to cellular scale without incrementing laser powers.

\begin{acknowledgement}

The authors thank Dr. Sunny Tiwari, Dr. Diptabrata Paul, and Dr. Shailendra Kumar Chaubey for valuable discussions related to this project. This work was partially funded by AOARD (grant number FA2386-23-1-4054) and the Swarnajayanti fellowship grant (DST/SJF/PSA-02/2017-18) to GVPK.

\end{acknowledgement}




\bibliography{achemso-demo}

\end{document}